\documentclass[pra,superscriptaddress]{revtex4-2}
\newcommand{\ket}[1]{\left | #1 \right\rangle}

\usepackage{epsfig}
\usepackage[utf8]{inputenc}
\usepackage[english]{babel}
\usepackage[T1]{fontenc}
\usepackage{amsmath}
\usepackage{amsthm}
\usepackage{amssymb}
\usepackage{bbm}
\usepackage{hyperref}
\usepackage{soul}

\usepackage[dvipsnames]{xcolor}

\begin{document}

\title{Valid and efficient entanglement verification with finite copies of a quantum state}

\author{Pawe{\l} Cie\'sli\'nski}
\email{pawel.cieslinski@phdstud.ug.edu.pl}
\affiliation{Institute of Theoretical Physics and Astrophysics, Faculty of Mathematics, Physics, and Informatics, University of Gdańsk, 80-308 Gdańsk, Poland}

\author{Jan Dziewior}
\affiliation{Max Planck Institute for Quantum Optics, 85748 Garching, Germany}
\affiliation{Faculty of Physics, Ludwig Maximilian University, 80799 Munich, Germany}
\affiliation{Munich Center for Quantum Science and Technology, 80799 Munich, Germany}

\author{Lukas Knips}
\affiliation{Max Planck Institute for Quantum Optics, 85748 Garching, Germany}
\affiliation{Faculty of Physics, Ludwig Maximilian University, 80799 Munich, Germany}
\affiliation{Munich Center for Quantum Science and Technology, 80799 Munich, Germany}

\author{Waldemar K{\l}obus}
\affiliation{Institute of Theoretical Physics and Astrophysics, Faculty of Mathematics, Physics, and Informatics, University of Gdańsk, 80-308 Gdańsk, Poland}

\author{Jasmin Meinecke}
\affiliation{Max Planck Institute for Quantum Optics, 85748 Garching, Germany}
\affiliation{Faculty of Physics, Ludwig Maximilian University, 80799 Munich, Germany}
\affiliation{Munich Center for Quantum Science and Technology, 80799 Munich, Germany}

\author{Tomasz Paterek}
\affiliation{Institute of Theoretical Physics and Astrophysics, Faculty of Mathematics, Physics, and Informatics, University of Gdańsk, 80-308 Gdańsk, Poland}
\affiliation{School of Mathematics and Physics, Xiamen University Malaysia, 43900 Sepang, Malaysia}

\author{Harald Weinfurter}
\affiliation{Institute of Theoretical Physics and Astrophysics, Faculty of Mathematics, Physics, and Informatics, University of Gdańsk, 80-308 Gdańsk, Poland}
\affiliation{Max Planck Institute for Quantum Optics, 85748 Garching, Germany}
\affiliation{Faculty of Physics, Ludwig Maximilian University, 80799 Munich, Germany}
\affiliation{Munich Center for Quantum Science and Technology, 80799 Munich, Germany}

\author{Wies{\l}aw Laskowski}
\email{wieslaw.laskowski@ug.edu.pl}
\affiliation{Institute of Theoretical Physics and Astrophysics, Faculty of Mathematics, Physics, and Informatics, University of Gdańsk, 80-308 Gdańsk, Poland}
\affiliation{International Centre for Theory of Quantum Technologies, University of Gdańsk, 80-308 Gdańsk, Poland}

\begin{abstract}
Detecting entanglement in multipartite quantum states is an inherently probabilistic process, typically with a few measured samples. The level of confidence in entanglement detection quantifies the scheme's validity via the probability that the signal comes from a separable state, offering a meaningful figure of merit for big datasets. Yet, with limited samples, avoiding experimental data misinterpretations requires considering not only the probabilities concerning separable states but also the probability that the signal came from an entangled state, i.e. the detection scheme's efficiency. We demonstrate this explicitly and apply a general method to optimize both the validity and the efficiency in small data sets providing examples using at most 20 state copies. The method is based on an analytical model of finite statistics effects on correlation functions which takes into account both a Frequentist as well as a Bayesian approach and is applicable to arbitrary entanglement witnesses.
\end{abstract}

\maketitle

\section*{Introduction}

Quantum entanglement is long recognized as an important prerequisite of modern quantum technologies.
Its detection is accordingly a well studied topic with a plethora of different methods available.
The field has evolved towards strategies directly applicable to experimental data which inevitably is limited to a finite number of detection events.
If this number is large, various forms of entanglement witnesses~\cite{Witness,Review, Guhne2009} provide reliable entanglement verification~\cite{finitedata2}.
Interestingly, also the analysis of smaller data sets allows to detect entanglement~\cite{finitedata}, most recently
with quantum state verification methods~\cite{qsv3,qsv, qsv2} and tailored game-like protocols in which particles are measured one by one~\cite{Bori1,Bori2,Bori3,Saggio2022}.
These methods are of high practical relevance in all cases where only a limited amount or only partial data is accessible, e.g.~when an inefficient source has to be characterized quickly or one is interested in entanglement detection in large-scale quantum systems.

\begin{figure}[!t]
\centering
\includegraphics[width=0.7\textwidth]{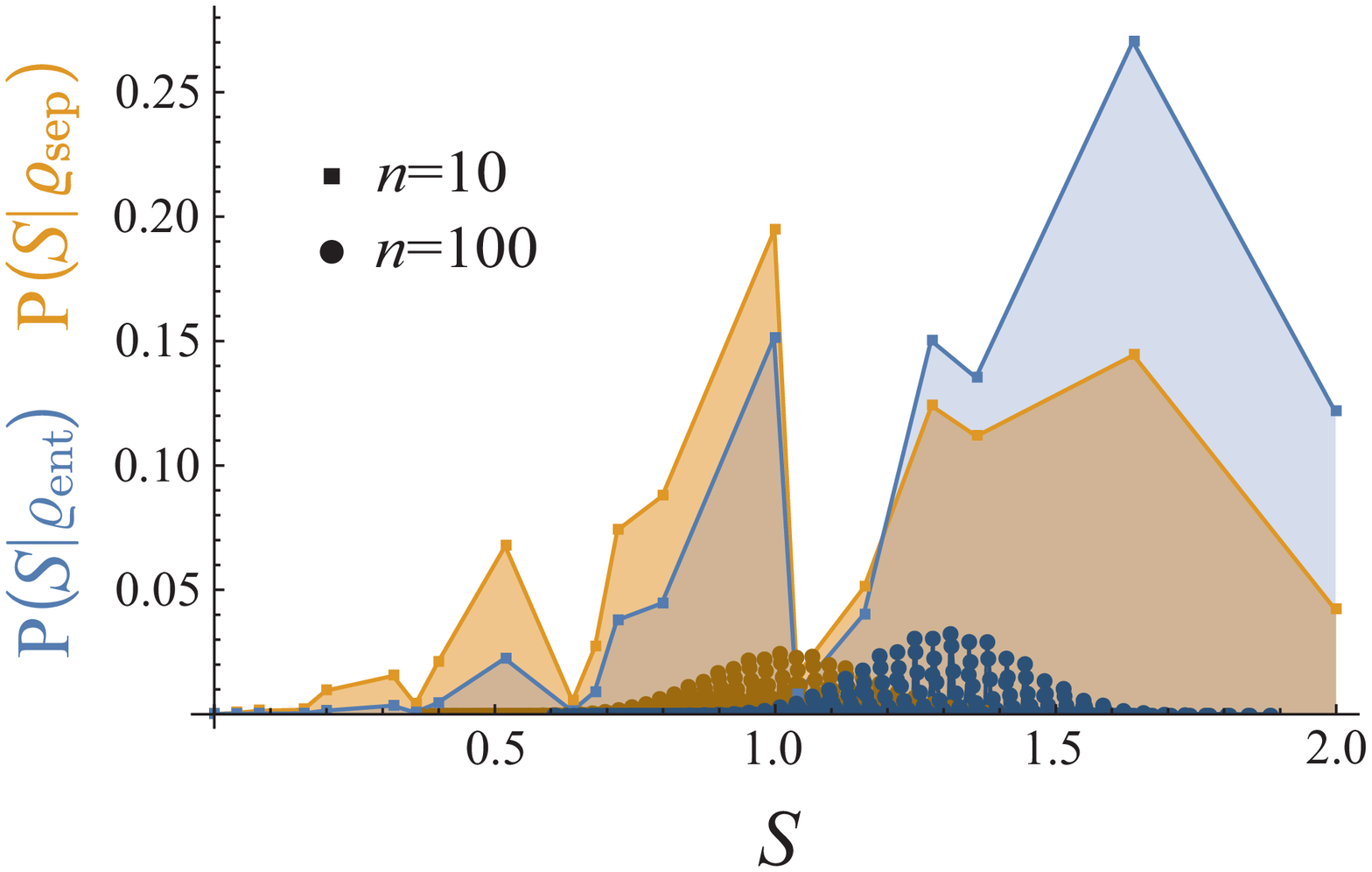}
\caption{\label{FIG1}
\textbf{Illustration of finite statistics effects on entanglement verification}.
The plot shows the probability distributions to obtain a specific value of a certain nonlinear entanglement witness $S$ discussed in the Methods (Certifying Entanglement). The witness uses
only two correlation measurements, each estimated either with $n=10$ (squares) or $n = 100$ (circles) copies of the states.
The probability distribution given an entangled state is marked in blue.
The probability in orange is computed for a separable state that saturates the value of the witness in the case of infinite data sets, i.e. a separable state that tends to yield large values of $S$.
For larger statistics (see circles), the two distributions have small overlap making it easy to verify that an entangled state was prepared.
For smaller statistics (see squares) the overlap is very significant and care has to be taken even if big values of the witness are observed. More detailed discussion on this issue is provided in the Results section (Limited Significance of Entanglement Certification).
Here we develop universal tools, applicable to any entanglement witness, that extend it to the domain of finite statistics and can be employed to efficiently use every single copy of a state. For examples see Optimising Entanglement Cerification Strategies in the Results section.
}
\end{figure}

The finiteness of data sets used to derive a conclusion about whether the state is entangled or not inevitably leads to a probabilistic nature of this conclusion.
Common certification schemes aim to ensure a small error if the outcome of the method speaks in favour of entanglement, i.e.~a small error of a false positive.
Yet, in the case of low statistics, this might come at the price of a large probability of rejecting an actually entangled state, i.e.~a false negative.
This tradeoff is illustrated in Fig.~{\ref{FIG1}}, which shows that for smaller statistics one is required to go to more and more strict criteria to reliably distinguish between results compatible with entangled states and results compatible with separable states.
In consequence, due to the increased strictness, it becomes also more and more improbable that an entangled state will pass the test.
It thus becomes crucial to properly apply the apparatus of statistical hypothesis testing~\cite{Lehmann2005,Robert2013,Berger2010} to the scenario of entanglement certification in order to avoid significant misinterpretations of conclusions based on small sample sizes and to properly optimize the usage of these scarce resources.

Here, we analyze the consequences of limited datasets for the application of entanglement witnesses based on correlation functions.
As mentioned above, there are competing interests for choosing a certification strategy. The first is achieving a certain level of validity, i.e.~trust in a positive result of the entanglement certification.
The second interest is to ensure that the test can correctly assess as many input states as possible, which we call its efficiency.
In general, different certification strategies will exhibit different efficiencies even if they achieve the same validity and vice versa.
Thus, both quantifiers should be taken into account when choosing an optimal test strategy.
We develop an algorithm for the optimization of experimental entanglement tests which simultaneously ensures high validity with optimal efficiency while being independent of the system's size.

Although the general framework of statistical analysis which is employed here is well known to statisticians, it has not yet been comprehensively applied in the field of entanglement certification, where it becomes especially relevant in the context of limited datasets.
The present work contributes to closing this interdisciplinary gap.
The results are illustrated with various examples where different entanglement witnesses are applied to general classes of quantum states.
While our examples all consider systems of qubits (the most commonly encountered systems in quantum information), our general analysis is by no means limited to two-dimensional systems and can be applied to any witness which is a function of generalized correlation tensor elements.

\section*{Results }
\label{SEC_MODEL}

\subsection*{Entanglement Certification as Hypothesis Testing}

There are two main approaches to statistical hypothesis testing, the Frequentist and the Bayesian one (see e.g.~\cite{Lehmann2005} for a Frequentist and~\cite{Robert2013,Berger2010} for a Bayesian perspective).
These approaches employ different quantifiers to capture different figures of merit.
In general, both are useful to statistically evaluate an entanglement certification scheme, but dependent on the context, such as e.g.~the availability of prior information or the practical goal of certification, one of these frameworks might be more suited than the other.
In the following we will first express entanglement detection as a hypothesis test and subsequently show how it is statistically analyzed from the Frequentist or Bayesian viewpoints.

Entanglement certification can be reduced to a so called simple hypothesis test, where the parameter space contains just two elements corresponding to the two disjoint hypotheses ‘the state is separable’ and ‘the state is entangled’.
We denote the two parameters as ‘sep’ and ‘ent’, respectively.
The outcome space $\mathcal{O}$ is given by the set of all possible outcomes generated by a particular entanglement detection function under consideration.
Depending on the chosen approach to statistical inference, the theory of hypothesis testing now allows to construct the optimal test to choose between the two hypotheses for any given outcome $Q \in \mathcal{O}$.

In non-randomized tests the outcome space is divided into two disjoint sets $\mathcal{O}_{\mathrm{acc}}$ (acceptance set) and $\mathcal{O}_{\mathrm{rej}} = \mathcal{O} \setminus \mathcal{O}_{\mathrm{acc}}$ (rejection set) such that entanglement is certified whenever $Q \in \mathcal{O}_{\mathrm{acc}}$.
Note, that sometimes it can be optimal to perform a (partially) randomized test, where there exists a subset of outcomes for which the decision between the hypotheses is made randomly. 
For simplicity, we present our analysis in terms of non-randomized tests and note that randomized test can be treated equivalently.
Furthermore, the usefulness of a randomized test is illustrated in the generalised scenario part of the Optimising Entanglement Certification Strategies section.

In typical entanglement certification schemes the outcomes $Q$ approximate theoretical quantities ${Q}^\infty$ which are functions of the states
that would be observed when the number of measurements tends to infinity. From this point on we use the superscript $\infty$ to indicate such ideal values.
The inference of entanglement is usually justified by the fact that values in the acceptance set $\mathcal{O}_{\mathrm{acc}}$ cannot be attained by any separable state, i.e.
\begin{align}
    \label{eq.certain}
    \mathrm{P}({Q}^\infty \in \mathcal{O}_{\mathrm{acc}}|\mathrm{sep}) = 0.
\end{align}
If the experiment is performed with sufficient statistical data the experimental result $Q$ approximates the theoretical $Q^{\infty}$ very well and thus for a result in $\mathcal{O}_{\mathrm{acc}}$ entanglement can be certified essentially with certainty.
Crucially, however, if the amount of empirical data is sufficiently small, the statistical fluctuations of the outcomes lead to a manifestly statistical nature of the conclusion and a proper analysis of the validity of the statistical inference becomes necessary.

\subsubsection*{Frequentist approach}

In the Frequentist approach, the validity of the test is given by the level of confidence $q_{\mathrm{F}} = 1 - \alpha$, that one or the other hypothesis is correct, with $\alpha$ being the level of significance.
In our simple certification scenario, the significance $\alpha$ is the upper bound on the probability $q_{\mathrm{F}}$ with which the experiment will wrongly certify entanglement if the state is separable.
As this is a statement about the average performance of the test, it does not allow any probabilistic assertions about the correctness of a particular single run of the procedure.
For example, if we perform certification of entanglement with the values typically used for hypothesis testing, i.e. $q_{\mathrm{F}}=95\% (\alpha=5\%)$, but are given separable states only, we would certify the state to be entangled on average in roughly $5\%$ of the runs.

In a certification scenario, the level of significance only depends on the likelihood of type I errors (false positives).
Thus, to achieve a certain validity $q_{\mathrm{F}}$, the acceptance set has to be chosen such that the likelihood of yielding a value in the acceptance set given a separable state is at most $1 - q_{\mathrm{F}}$, i.e.,
\begin{align}
    \label{eq.freq.valid}
    \sum_{Q \in \mathcal{O}_{\mathrm{rej}}} \mathrm{P}(Q|\mathrm{sep}) \geq q_{\mathrm{F}}.
\end{align}
The likelihood function for the opposite hypothesis $\mathrm{P}(Q|\mathrm{ent})$ is irrelevant as far as the level of significance is concerned.

\begin{figure}[!t]
\centering
\includegraphics[width=\textwidth]{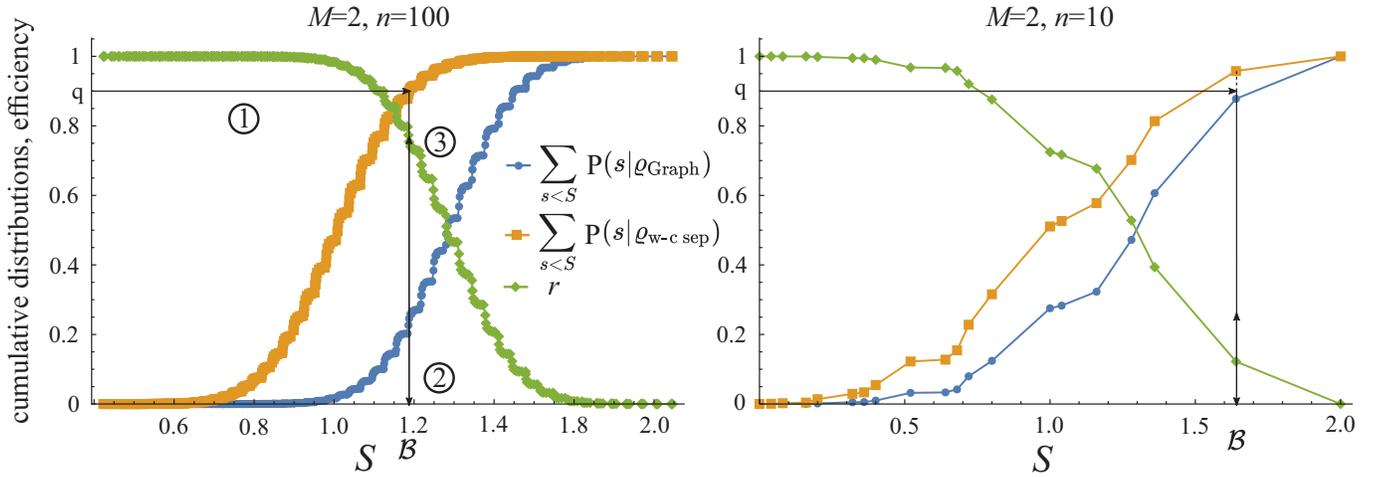}
\caption{\label{FIG_CUMUL}
\textbf{Validity and power in the Frequentist approach applied to the nonlinear witness $S$.}
Given a required minimal, e.g., $q_{\mathrm{F}} =0.9$, an appropriate outcome space over the values of $S$ (for definition see Entanglement Certification in Metods) is chosen by specifying a bound $\mathcal{B}$ such that the separable correlations have at least $q_{\mathrm{F}}$ mass of its probability distribution below $\mathcal{B}$, i.e., the cumulative probability distribution of the worst case separable correlations (orange data) reaches $q_{\mathrm{F}}$ (arrows marked with encircled 1 and 2).
For this bound, one then evaluates the cumulative probability of the entangled state (blue data). 
The power, $r$, is defined as one minus this probability and is plotted in green.
The power at the determined bound is marked by an arrow with encircled 3.
The higher the power, the better the entanglement detection works, i.e., the higher the probability to actually detect the entangled state as such, see Fig.~\ref{FIG1}. 
Reducing the statistics from $n=100$ (left panel) to $n=10$ (right panel) lowers the power of the detection from about $75\%$ to about only $12\%$ if the same validity $q_{\mathrm{F}} = 0.9$ (level of significance $10\%$) is required.
}
\end{figure}

Although the distinction between types of errors in Frequentist inference is well known, in particular in the field of entanglement certification with small data sets, often the focus lies exclusively on the level of significance~\cite{Bori1,Bori2,Bori3, finitedata, Jungnitsch2010, Minh1}.
We want to argue, that in most practical applications, one is not only interested in a valid method, but also in its ability to positively certify at least some entangled states.
This tradeoff becomes particularly crucial in the case of very low statistics.
Thus, hypothesis tests should also be
optimized based on a second quantifier, the power of the test, 
which characterizes exactly this ability to provide a positive certification corresponding to minimizing $e_2$, the error of type II (probability of false negatives).
The power $r$ is given by
\begin{align}
\label{eq_freq_eff}
r = 1-e_2 = \sum_{Q \in \mathcal{O}_{\mathrm{acc}}} \mathrm{P}(Q|\mathrm{ent}).
\end{align}
The concept of statistical power is essential for finding optimal tests and to properly assess the usefulness of statistical conclusions.
See Fig.~\ref{FIG_CUMUL} for an illustration.
Thus, in our terminology, the maximization of power corresponds to a maximization of the efficiency of the test.

\subsubsection*{Bayesian approach}

In a Bayesian approach to certification, probabilities for both hypotheses can be calculated given a certain outcome, $Q$, i.e.~$\mathrm{P}(\mathrm{ent}|Q)$ and $\mathrm{P}(\mathrm{sep}|Q) = 1 - \mathrm{P}(\mathrm{ent}|Q)$.
Validity in this context is readily identified with the Bayesian level of acceptance $q_{\mathrm{B}}$, which is the minimal value of the posterior probability for which the hypothesis $\mathrm{ent}$ is accepted.
Entanglement is thus certified whenever $\mathrm{P}(\mathrm{ent}|Q) \geq q_{\mathrm{B}}$.
This means that the acceptance set $\mathcal{O}_{\mathrm{acc}}$ has to be chosen such that
\begin{align}
    \label{eq.bayes.valid}
    \mathrm{P}(\mathrm{ent}|Q) \geq q_{\mathrm{B}}, \quad \textrm{ for all } \quad Q \in \mathcal{O}_{\mathrm{acc}},
\end{align}
i.e., whenever a value of $Q \in \mathcal{O}_{\mathrm{acc}}$ is observed, the certification of entanglement is credible at least with degree $q_{\mathrm{B}}$.
Note, that as given by Bayes theorem, the posterior $\mathrm{P}(\mathrm{ent}|Q)$ is a function of both $\mathrm{P}(Q|\mathrm{sep})$ and $\mathrm{P}(Q|\mathrm{ent})$, as well as of a the quotient of priors $\mathrm{P}(\mathrm{ent}) / \mathrm{P}(\mathrm{sep})$.

To arrive at a formulation of Bayesian hypothesis testing in terms of validity and efficiency, one has to consider the connection between the level of acceptance and so called loss functions.
For a given outcome $Q$ and weights $\beta$ and $(1-\beta)$ for false positive and false negative respectively, the (posterior) loss function $L_Q$ is given by
\begin{align}
    L_Q = \begin{cases}
        \beta \; \mathrm{P}(\mathrm{sep}|Q) & \text{if } Q \in \mathcal{O}_{\mathrm{acc}} \\
        (1-\beta) \; \mathrm{P}(\mathrm{ent}|Q) & \text{if } Q \in \mathcal{O}_{\mathrm{rej}},
    \end{cases}
\end{align}
i.e.~the respective probability of error multiplied by the corresponding weight.
It can be shown that for each $Q$ this loss is optimized if the rejection and acceptance sets are chosen simply based on an acceptance level of $q_{\mathrm{B}} = \beta$~\cite{Robert2013}.
Thus for optimal tests, we can simply identify the acceptance level $q_{\mathrm{B}}$ with the (relative) weight $\beta$ given to the error of false positive.

A priori, i.e., without the knowledge of the outcome $Q$, an expected loss $L$ can be defined as
\begin{align}
    \label{eq.bayes.loss}
    L = \sum_Q L_Q \mathrm{P}(Q) = q_{\mathrm{B}} \mathrm{P}(Q \in \mathcal{O}_{\mathrm{acc}}|\mathrm{sep}) \mathrm{P}(\mathrm{sep}) + (1-q_{\mathrm{B}}) \mathrm{P}(Q \in \mathcal{O}_{\mathrm{rej}}|\mathrm{ent}) \mathrm{P}(\mathrm{ent}),
\end{align}
which corresponds to a weighted sum of the probabilities for the errors of both types.
This is also equivalent to the average loss if the procedure is applied multiple times.
In analogy to the power for Frequentist tests, the average loss $L$ allows to compare different Bayesian test strategies with equal acceptance levels.
From a Bayesian perspective maximal efficiency of a test thus corresponds to minimal loss $L$.

Finally, let us also briefly mention why the relevant priors are not discussed in the usual detection of quantum entanglement, via witnesses, in the limit of many measurement results~\cite{finitedata2}.
In such a case, when the number of state copies is large: $n \gg 1$, it is possible to choose acceptance sets $\mathcal{O}_{\mathrm{acc}}$ for which it is practically impossible for a separable state to exceed the bound of the witness, i.e., $\mathrm{P}(Q \in \mathcal{O}_{\mathrm{acc}} | \mathrm{sep}) \approx 0$, as discussed above.
Therefore, Bayes rule simplifies and the posterior becomes $\mathrm{P}(\mathrm{ent} | Q) \approx 1$ independently of the priors.
In consequence, for any outcome in the acceptance set entanglement is certified with validities $q_{\mathrm{F}} \approx q_{\mathrm{B}} \approx 1$ both from the Frequentist and the Bayesian perspective
 and the distinction between the two approaches becomes superfluous.

\subsection*{Effects of Finite Statistics}

The standard experimental scheme for entanglement certification in qubit systems is based on the measurement of correlation functions or simply correlations $T_{\mu_1 \dots \mu_N}$ defined by
\begin{align}
T_{\mu_1 \dots \mu_N } = \langle R_{\mu_1} \ldots R_{\mu_N} \rangle = \mathrm{Tr}(\rho \, \sigma_{\mu_1} \otimes \dots \otimes \sigma_{\mu_N}),
\label{EQ_T}
\end{align}
where $\sigma_{\mu_k}$ is the $\mu_k$-th local Pauli matrix of the $k$-th party with $\mu_k \in \{0,x,y,z\}$ and $\sigma_0$ being the identity.
The correlation is thus the average of the product of local measurement outcomes $R_{\mu_k} = \pm 1$ when suitable local Pauli measurements are executed.
Note that definition~(\ref{EQ_T}) does not only include the full correlations, for which all parties perform measurements but also marginal correlations, i.e.~values of $T$ to which only a subset of parties contributes, which formally correspond to setting the remaining indices $\mu_k = 0$.
To simplify the notation in the following we write the correlations as $T_j$, where the index $j \in \left[1, 4^N\right]$ indicates the set of chosen measurement settings over all parties.

Consider a broad class of procedures for entanglement certification, in which a quantity $Q^{\infty}$, e.g., an entanglement witness, is calculated as a function of the correlations $T_j$.
Note that by choosing $Q^{\infty}$ tailored to witness various types of entanglement, e.g. genuine multipartite entanglement, one can employ the further discussed method for testing the statements about, e.g., genuine multipartite entanglement or biseparability.
For the usual application of the conditions imposed by $Q^{\infty}$, the actual correlation functions are estimated with a finite number of experimental trials. 
We denote the actually measured correlations resulting from finite statistics by $\tau_j$.

Thus, instead of the ideal quantum mechanically predicted value $Q^\infty$ based on $T_j$ obtained in the limit of infinite statistics, experimentally we have only access to the modified value $Q$, calculated from $\tau_j$. We therefore begin with modelling $\tau_j$.

Let $\{n_j\}$ denote the number of state copies measured for each setting $j$, which can vary from setting to setting. For the $j$th measurement setting we have:
\begin{equation}
\tau_j = \frac{n^+_j - n^-_j}{n_j}
\label{EQ_TAU}
\end{equation}
where $n^{\pm}_j$ are the numbers of trials in which the product of local results equals $R_{\mu_1} \cdots R_{\mu_N} = \pm 1$, with $n^+_j + n^-_j = n_j$.

For large $n_j$ we obtain $\tau_j \approx T_j$, but for small $n_j$ the measured $\tau_j$ take potentially different values which are furthermore restricted to the discrete set $\tau_j \in \{ -1, -1 + \frac{2}{n_j}, \dots, 1 - \frac{2}{n_j}, 1\}$.
Given an $N$-qubit state, expressed by its correlations $T_j$, it is possible to explicitly calculate the ideal probability distribution $\mathrm{P}(\tau_j)$ over these discrete values by employing the binomial distribution as demonstrated in Methods (Witnesses in finite statistics).
The mean and variance of this distribution are given as
\begin{equation}
\langle \tau_j \rangle = T_j, \qquad \mathrm{Var}(\tau_j) = \sigma_j^2/n_j,
\end{equation}
where $\sigma_j^2$ is a function of the ideal quantum prediction equal to $\sigma_j^2 = 1 - T_j^2$, see also \cite{KnipsFiniteStatistics}.
Note that the variance depends on the correlation value itself, e.g.~a perfect correlation has no variance.
This already hints that the method is most useful for states with high values of $T_j^2$.
 
Considering a particular entanglement certification procedure with the outcome $Q=f(\tau_{j_1},\ldots,\tau_{j_M})$ ($M$ being the number of correlation measurements),
it becomes possible to calculate explicitly the probability distribution over the outcomes
\begin{equation}
\mathrm{P}(Q)=\sum_{\substack{\tau_{j_1},\ldots,\tau_{j_M} \\ f(\tau_{j_1},\ldots,\tau_{j_M})=Q}} \, \prod_{k=1}^M \mathrm{P}(\tau_{j_k}).
\end{equation}
In the Methods, Witnesses in Finite Statistics section, we illustrate this procedure for both linear ~\cite{Witness, Review,Guhne2009} and non-linear entanglement witnesses~\cite{Badziag,HJ1,HJ2,HJ3,Minh1,Minh2,Knips2016}. The linear witness $\mathrm{W} = \sum_j \alpha_j \sigma_j$, with expectation value $\mathrm{Tr}(\mathrm{W} \rho) = \sum_j \alpha_j T_j$ denoted as $E_{\mathrm{W}}^{\infty}$, extension to finite statistics has the following properties:
\begin{align}
 \langle E_{\mathrm{W}} \rangle  = E_{\mathrm{W}}^\infty, \quad \mathrm{ and } \quad \mathrm{Var}(E_{\mathrm{W}}) = \sum_{j} \alpha_j^2 \mathrm{Var}(\tau_j).
\label{EQ_Efin_meanVar}
\end{align}
For nonlinear witnesses, further work is required as one needs to calculate the effect of finite statistics on nonlinear functions of $\tau_j$.
We describe a general procedure in the Methods and here briefly illustrate the case when the witness depends on the square of correlations.
Its mean and variance read
\begin{equation}
\langle \tau_j^2 \rangle = T_j^2 + \mathrm{Var}(\tau_j), \qquad \mathrm{Var}(\tau_j^2) = \frac{2(n_j-1)(1-T_j^2)[(2n_j-3) T_j^2 + 1]}{n_j^3}.
\label{EQ_SQUARE}
\end{equation}
Note that the mean value of the square has a consistent shift by exactly the variance of the estimation from the finite data.
This is a consequence of taking the square --- the negative values in the distribution are transformed into positive values leading to a systematic deviation.
This fact is important in the experimental analysis of finite data sets and has already been addressed in Ref.~\cite{Knips}.
As a final illustration assume the quadratic witness $S^{\infty}=\sum_j^M T_j^2$ that sums up 
$M$ squared correlations (see Certyfing Entanglement in Methods), each measured with the same number of copies $n$, i.e. $n_j = n$. We obtain
\begin{equation}
\langle S\rangle = \frac{(n-1)S^\infty_M+M}{n}, \quad \mathrm{with} \quad \mathrm{Var}( S ) = \sum_{i = 1}^M \mathrm{Var}(\tau_i^2),
\label{EQ_Sfin_meanVar}
\end{equation}
where now apart from the statistical uncertainty also a shift of the mean value is present.

\subsection*{Limited Significance of Entanglement Certification}
\label{SEC_MODEL_LimSig}

The modelling of probability distributions for the outcomes of single correlation measurements now allows to calculate explicitly the probability distribution over outcomes of the entanglement witnesses $Q$.
This calculation is always based on particular state spaces corresponding to the hypotheses $\mathrm{ent}$ and $\mathrm{sep}$.
Initially, for simplicity we consider a simple scenario, where the test is supposed to distinguish between having a particular entangled state of the form
\begin{align}
    \rho = \frac{3}{4} |\mathrm{GHZ}\rangle \langle \mathrm{GHZ} | + \frac{1}{4} \, \openone/2^N,
    \label{eq_rho}
\end{align}
or an arbitrary separable state.
Here, $\ket{\mathrm{GHZ}}$ is an $N$-qubit GHZ state, for any $N$, and $\openone/2^N$ corresponds to the so-called white noise with no correlations at all.
Note that in realistic scenarios the expected distribution of input states might be more complicated as we also demonstrate below.
The probability distributions corresponding to the $\mathrm{sep}$ hypothesis are modeled based on a worst case estimation of correlations compatible with separability as discussed in detail in Methods (Estimating Probability Distributions \& Correlations compatible with separability).

Now consider the linear witness $\mathrm{W}=\sigma_y \otimes \sigma_y \otimes \sigma_x \otimes \cdots \otimes \sigma_x - \sigma_x  \otimes \cdots \otimes \sigma_x + 1$ with only $n = 10$ clicks per single correlation measurement, i.e., 20 state copies in total.
Due to this very small number of data points separable states can now give rise to outcomes in the acceptance set, i.e., $\mathrm{P}(E_{\mathrm{W}}<0|\mathrm{sep}) > 0$, where we follow the convention that separable states admit $E_{\mathrm{W}}^{\infty} \ge 0$ in the case of infinite statistics.
In fact, in this example the probability that the witness is negative on a suitable separable state is as large as $41.5\%$.
The straightforward solution seems to be to make the boundary more strict, i.e.~require sufficiently negative values to certify entanglement.
However, even with the strictest nontrivial bound of $-4/5$ we find that the violation by a separable state is still equal to $2.5\%$, i.e.~$\mathrm{P}(E_{\mathrm{W}} \le -4/5 | \mathrm{sep}) = 2.5\%$.
At the same time, the state $\rho$ gives rise to values  $E_{\mathrm{W}} < -4/5$ with probability $26.7\%$.
While it is clearly more probable for the entangled state to violate the bound than for any separable state, still, given the finite values of both probabilities, the question remains what the appropriate probabilistic conclusion about the quality of such certification should be.

The nonlinear criterion $S=\langle \sigma_y \otimes \sigma_y \otimes \sigma_x \otimes \cdots \otimes \sigma_x \rangle^2 + \langle\sigma_x  \otimes \cdots \otimes \sigma_x \rangle^2$ has been illustrated in Fig.~\ref{FIG1}.
Again we consider the sum of only two squared correlations, each estimated in $10$ experimental trials.
Our analytical tools return the plotted $\mathrm{P}(S|\mathrm{ent})$ based on the above mentioned entangled state $\rho$.
Fig.~\ref{FIG1} clearly illustrates that finite statistics leads to the violation of the bound of $1$ by a separable state.
Again the immediate solution seems to be making the bound more strict.
We find that $\mathrm{P}(S = 2|\mathrm{sep}) = 4.2\%$, but at the same time $\mathrm{P}(S = 2|\rho) = 6.9\%$.
With this bound the procedure would wrongly certify entanglement (false positive) in at most only $4.2\%$ of the cases, but at the same time, it would fail to correctly certify it (false negative) in roughly $93\%$ of the cases where the state was entangled.
Furthermore, for any single experiment yielding value $S = 2$, we are clearly not justified to conclude that the state is entangled, as a separable state might have given rise to the same value of $S$ with a comparable probability.

This type of explicit calculation can be applied to any witness based on correlations values to evaluate whether it can be applied efficiently in the case of finite statistics.
In particular, it becomes necessary to put $\mathrm{P}(Q \in \mathcal{O}_{\mathrm{acc}}|\mathrm{sep})$
in relation to the probability $\mathrm{P}(Q \in \mathcal{O}_{\mathrm{acc}}|\mathrm{ent})$ via a proper statistical analysis, either Frequentist or Bayesian.

\subsection*{Resource Optimization}
\label{SEC_OPT}

Based on the possibility to quantitatively evaluate any entanglement certification procedure in terms of validity and efficiency, we now tackle the question how to best use a small number $\eta$ of available copies of a multi-qubit quantum state.
The experimental choice to be made is about the amount of settings $M$ and the distribution of the copies among those $\{n_j\}$, where $\sum_j n_j = \eta$.
Furthermore, a suitable outcome space $\mathcal{O}_{\mathrm{acc}}$ has to be chosen.
By explicitly calculating the distributions $\mathrm{P}(Q|\mathrm{sep})$ and $\mathrm{P}(Q|\mathrm{ent})$, and assuming a certain prior the validity and efficiency of the procedure can be characterized, no matter if a Frequentist or Bayesian approach is chosen.
In practice a reasonable scenario could be to require a certain minimal validity and then choose a combination of experimental parameters with a suitable acceptance space, such as to maximize the efficiency.
This yields the following optimization problem:
\begin{quote}
    Given a certain minimal validity $q_\mathrm{min}$ and a total number $\eta$ of state copies, what set of $M$ observables, each estimated on $\{n_j\}$ copies of the state, and what acceptance set $\mathcal{O}_{\mathrm{acc}}$ are to be chosen such that an observation of $Q \in \mathcal{O}_{\mathrm{acc}}$ certifies entanglement with a validity $q \geq q_\mathrm{min}$, while keeping the efficiency as high as possible.
\end{quote}

Note that when $Q$ is a function of marginal correlations corresponding to commuting operators it can be possible to measure them simultaneously from a single global measurement setting.
This is the approach chosen in \cite{Bori1}, where a single state copy is used to obtain values for several marginal correlations  providing the outcome for a linear witness.
Thus, in general, given a function $Q$ also of marginal correlations it is of high interest to group these into as few measurement settings as possible.
Since our example witnesses $S$ and $W$ are solely based on full correlations this particular layer of optimization is not relevant for them.

\subsection*{Optimising Entanglement Certification Strategies}
\label{SEC_EXAMPLES}
Scenario: First of all, let us emphasise that the proper statistical analysis is especially crucial in the case of a very limited data set composed of at most a few tens of data points.
In the examples below we assume that no more than $20$ copies of multi-particle state from a possibly entangled source have been successfully measured.
The small amount of data makes it practically impossible to distinguish entangled states which are close to separable ones. 
Thus, the best candidates are states which are strongly non-classically correlated.
Generalizing Eq.~(\ref{eq_rho}), here, we therefore consider mixed states of the form
\begin{equation}
\rho = p |\psi \rangle \langle \psi | + (1-p) \rho_{\mathrm{n}},
\label{EQ_EX_STATE}
\end{equation}
where a dominant admixture comes from a pure state $\ket{\psi}$, assumed to admit perfect correlations, and the remaining part, $\rho_N$, describes experimental noise.
Since our theoretical analysis just depends on the number of correlations (measurement settings) it is thus independent of the number of involved particles. 
See Ref.~\cite{Flammia2011} for fidelity estimation with the same feature.
Accordingly, we will not specify the state $\ket{\psi}$ exactly as the only parameters that enter the witnesses are correlations. 
We would like to emphasise this point, the methods work for any GHZ states, graph states, cluster states, absolutely maximally entangled states etc.
Although in practice the perfect correlations of $\ket{\psi}$ will be reduced by various types of noise, which we consider via the addition of $\rho_n$, we show in Methods (Effects of different types of noise) that the probability to observe a certain measurement outcome depends almost entirely on the parameter $p$, and not a type of noise as long as it is uniform and unbiased.
Clearly, it is possible to introduce a specific noise that results in different success probability characteristics.
In such a case one can still apply our methods, but here we focus on random noises as they are the usual models of experimental imperfections.
Thus, we restrict our examples to the case of white noise, i.e.~$\rho_{\mathrm{n}} = \openone/2^N$.
In effect, the model (\ref{EQ_EX_STATE}) encompasses a plethora of experimental situations that are of interest in quantum information. 

Consider a certification scenario where we expect that the tested state is either from the family (\ref{EQ_EX_STATE}) or a separable state modelled according to the worst-case correlations compatible with separability (see Estimating Probability Distributions and Correlations Compatible with Separability in the Methods), with $\mathrm{P}(\mathrm{sep}) = \mathrm{P}(\mathrm{ent}) = 1/2$.
We assume 20 available state copies and require a target validity of $q_{\mathrm{F}}=q_{\mathrm{B}}=97.5\%$.
This will help us to compare and highlight the differences in reasoning that occur from Frequentist and Bayesian perspectives on statistical inference.
According to our optimization scheme, we want to determine the set of measurements, i.e., their amount $M$ and in general also the measurement directions, together with an optimal distribution of clicks per setting $\{n_j\}$. 
For simplicity here, we restrict the presentation to cases where we assume that all $n_j$'s are the same and equal to $n$, and hence we constrain the product $M n = 20$. 
Let us first set $p=3/4$  and specify the Bayesian loss function as (\ref{eq.bayes.loss.est}). 

\begin{figure}[!t]
\centering
\includegraphics[width=1\textwidth]{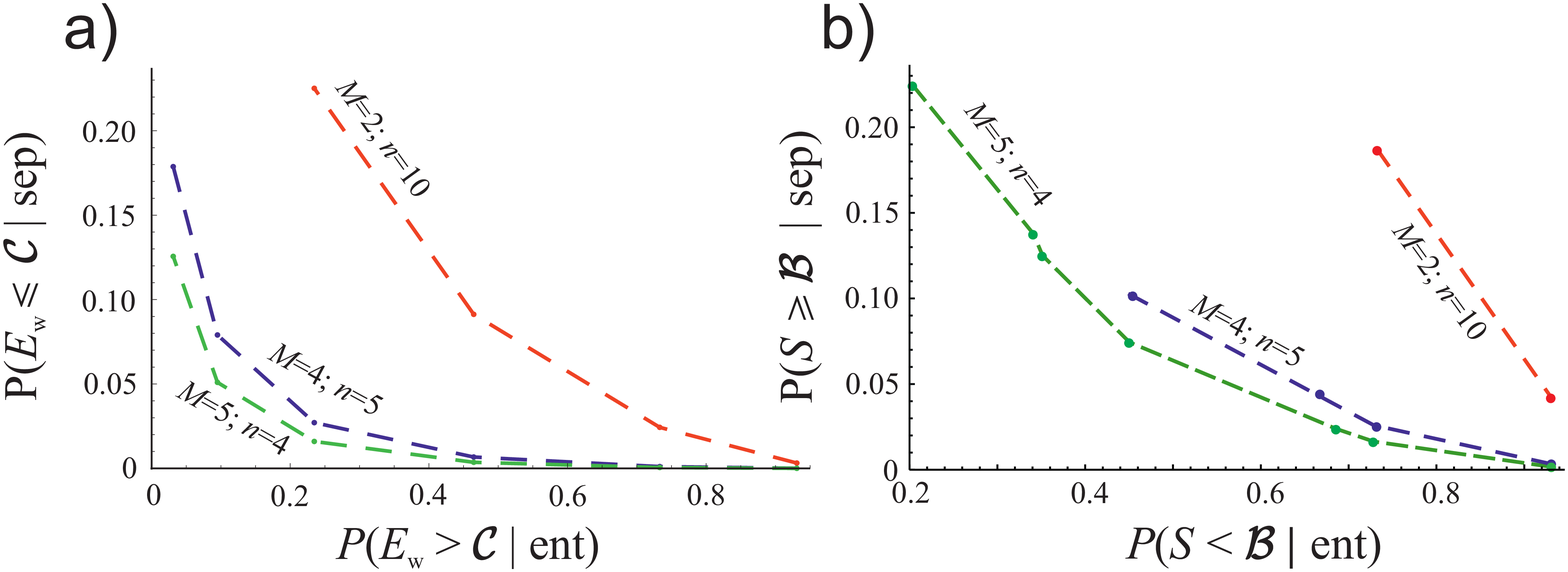}
\caption{\label{FIG_20OPT}
\textbf{Optimal use of 20 copies.} Plot for $\rho$ in Eq.~(\ref{EQ_EX_STATE}) with $p = 3/4$, e.g. a GHZ state mixed with white noise.
Panel a) is for the linear witness, panel b) for the quadratic witness.
The discreteness of the outcome space for just $20$ copies is clearly visible in the plot.
To maximize validity and efficiency both probabilities need to be simultaneously minimized. 
This is best realised by dividing the 20 copies into 5 correlation measurements, each correlation estimated in 4 experimental runs.
}
\end{figure}

Linear witness: To illustrate how to optimally distribute the 20 copies for the linear witness $W$, as an example we choose a procedure where the acceptance set is chosen via an upper bound $\mathcal{C}$.
We extend the linear witness to $M$ different settings as $E_{\mathrm{W}}^{\infty}=T_1-T_2-\cdots -T_M+1$.
It turns out that the worst case separable state, i.e.~the state which maximizes the probabilities in the acceptance set, for a sufficiently low bound $\mathcal{C}$ has the correlations $T_1=-1/M$ and $T_i=1/M$ for $i=2,\ldots, M$.
In Fig.~\ref{FIG_20OPT}a) we plot $\mathrm{P}(E_{\mathrm{W}} \leq \mathcal{C}|\mathrm{sep})$ and $\mathrm{P}(E_{\mathrm{W}} > \mathcal{C}|\mathrm{ent})$, which are directly related to validity and efficiency.
As the simultaneous optimization of the two quantifiers is based on keeping both probabilities as small as possible, clearly the optimal procedure is realized for $M=5$ and $n=4$.

From the Frequentist perspective to achieve the minimally required validity of $97.5\%$, a bound of $\mathcal{C}_{\mathrm{F}}=-2.5$ has to be chosen, which leads to an efficiency of $r\approx 76.5\%$. The highest validity achievable with 20 state copies is $99.996\%$ with a bound of $\mathcal{C}_{\mathrm{F}} = -4$.
However, it comes at the cost of efficiency of just $r\approx 6.9\%$.

The Bayesian framework requires $\mathrm{P}(\mathrm{ent}|E_{\mathrm{W}}) \geq 97.5\%$ for all values in the acceptance set. This is satisfied if the bound $\mathcal{C}_{\mathrm{B}}$ is chosen as $-3$. The associated efficiency, in this case, is given by the loss function of $L\approx 0.007$.

From the above results, one can see a clear difference in the choice of the acceptance sets for each reasoning.
Namely, the Frequentist bound is less strict than the Bayesian counterpart, implying instances in which experimenters would disagree on the conclusions about the entanglement certification.
For comparison, from the Frequentist perspective the adjusted bound of $\mathcal{C}_{\mathrm{B}}=-3$ that is optimal for the Bayesian approach results in the level of confidence given by $99.64 \%$ and the corresponding power of $r \approx 53.5\% $.
As expected, for the lower bound confidence is increased and power reduced, however, we emphasise that in general the Bayesian approach does not always yield stricter acceptance sets.

Quadratic witness: Just as for the case of the linear witness, we introduce a bound $\mathcal{B}$ to define the acceptance set also in the example of the quadratic witness.
Fig.~\ref{FIG_20OPT}b) presents the two relevant quantities $\mathrm{P}(S\geq\mathcal{B}|\mathrm{sep})$ and $\mathrm{P}(S< \mathcal{B}|\mathrm{ent})$. 
Again, the best use of copies is given by $n=4$ and $M=5$.

Based on the minimally required validity of $97.5\%$, we calculate the bound $\mathcal{B}$ for the Frequentist approach to be $\mathcal{B}_{\mathrm{F}}=4$ and in the Bayesian approach we get $\mathcal{B}_{\mathrm{B}}=5$.
This leads to an efficiency of $r \approx 31.4\%$ and $L \approx 0.013$ respectively.
Recall that the chosen priors for the Bayesian inference are $\mathrm{P}(\mathrm{ent})=\mathrm{P}(\mathrm{sep})=0.5$.
In general, the examined state is entangled iff $p > 1/(2^{N-1}+1)$~\cite{Pittenger2003}.
Thus, given our state model,
a more realistic scenario would take into account the cardinality of the separable states within the family (\ref{EQ_EX_STATE}) in the prior probabilities.
For example, with a uniform distribution over $p$ and $N=4$, the natural prior is $\mathrm{P}(\mathrm{ent})=8/9$ and $\mathrm{P}(\mathrm{sep})=1/9$.
This leads to the change of the acceptance set with $\mathcal{B}_{\mathrm{B}}=4$ and efficiency $L \approx 0.018$.
With increasing $N$ the efficiency in the Bayesian approach is changing.
As the last example, for $N=5$ the prior is $\mathrm{P}(\mathrm{ent})=16/17$ and $\mathrm{P}(\mathrm{sep})=1/17$,
the Bayesian bound is given by  $\mathcal{B}_{\mathrm{B}} = 3.5$ and the corresponding efficiency reads  $L \approx 0.015$.

Again, for equal priors Frequentist reasoning yields a greater acceptance set.
However, here one can explicitly see that for more physically motivated priors this is no longer true and $\mathcal{C}_{\mathrm{B}}$ can be smaller than $\mathcal{C}_{\mathrm{F}}$.
For all of the discussed cases, confidence and power are $(\mathcal{B}=5,q_{\mathrm{F}}=99.8\%,r=6.9\%)$, $(\mathcal{B}=4,q_{\mathrm{F}}=97.6\%,r=31.4\%)$ and $(\mathcal{B}=3.5,q_{\mathrm{F}}=92.5\%,r=54.9\%)$ respectively. Note that the last level of confidence does not meet the required validity condition.

Generalised scenario: Finally, in a more general example, we relax the condition of an equal amount of copies per setting and apply our optimization algorithm to determine the optimal amount of settings and distribution of state copies over these settings, such as to achieve optimal efficiency for certification with the quadratic witness $S$.
It turns out that different certification strategies are found to be optimal, depending on whether a Frequentist or Bayesian approach is chosen.
We calculate the distribution $\mathrm{P}(\rho_j|\mathrm{ent})$ based on the model of a family of noisy three-qubit GHZ states (\ref{eq_rho}), with a mean $\bar{p} = 0.8$ and a standard deviation $\Delta p = 0.1$, such that for the entangled states
\begin{align}
    \mathrm{P}(p) = \begin{cases}
  \mathcal{N} \mathrm{e}^{-\frac{(p-\bar{p})^2}{2(\Delta p)^2}}& p \in [p_\mathrm{min},1] \\
  0 & \text{else},
\end{cases}
\end{align}
with proper normalization $\mathcal{N}$ and the lower bound $p_\mathrm{min} = 1/5$ given by the requirement that the states have to be entangled, see the quadratic witness example in the previous subsection.
We set the maximal number of settings to $M=3$, allow at most $13$ state copies and require a minimal validity of $70\%$.
Furthermore, we again choose not to model the separable state space and instead estimate $\mathrm{P}(S|\mathrm{sep})$ via the worst case optimization used above.
In this example for the prior we assume that it is twice as likely to encounter an entangled state from our family than any separable state.

\begin{figure}[!t]
\centering
\includegraphics[width=0.99\textwidth]{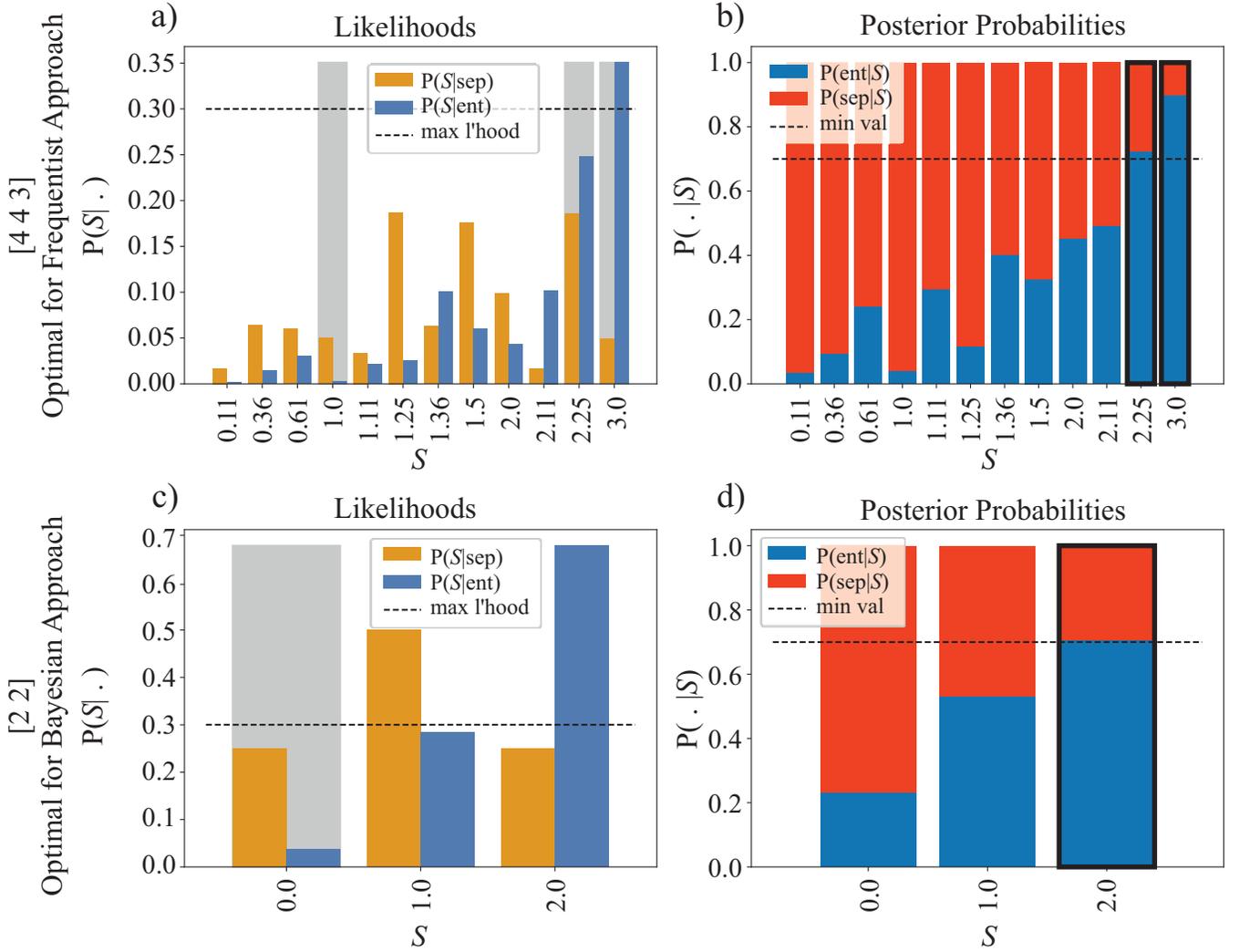}
\caption{
\textbf{Optimal choices of distributions of states for Bayesian and Frequentist figures of merit.}
Panel a) shows the estimated distributions $\mathrm{P}(S|\mathrm{ent})$ (blue bars) and $\mathrm{P}(S|\mathrm{sep})$ (orange bars) for the choice $[4~4~4]$ which is optimal from the Frequentist point of view.
Note that the shown $\mathrm{P}(S|\mathrm{sep})$ is actually the worst case estimate $\mathrm{G}_{\mathrm{acc}}(Q)$, optimized for one particular acceptance set.
The acceptance set $\mathcal{O}_{\mathrm{acc}} = \{0,1,2.25,3\}$ (marked by the gray background) is chosen by maximizing $\mathrm{P}(S \in {O}_{\mathrm{acc}}|\mathrm{ent})$, i.e., includes as much as possible of blue bar length while not allowing $\mathrm{P}(S \in {O}_{\mathrm{acc}}|\mathrm{sep})$ (estimated by $\mathrm{G}_{\mathrm{acc}}(S \in {O}_{\mathrm{acc}}|\mathrm{sep})$) to exceed $1 - 70\% = 30\%$ (dashed black line), i.e., not getting too much orange bar length into the acceptance set.
A power of $66.4\%$ is reached.
Panel b) shows the posterior probability $\mathrm{P}(\mathrm{ent}|S)$ for that choice of experiment.
The acceptance set (marked by the black boxes) is given by all values of $S$ for which $\mathrm{P}(\mathrm{ent}|S) = 1 - \mathrm{P}(\mathrm{sep}|S) \geq 70\%$ (dashed horizontal line) leading to $\mathcal{O}_{\mathrm{acc}} = \{2.25,3\}$.
Note, that the posterior is based on the point-wise worst case optimization $\mathrm{G}_S$ which is different from the one employed in panel a), see Estimating Probability Distributions in the Methods.
Panels c) and d) show the respective distributions for the choice $[5~3~3]$ optimal from the Bayesian point of view.
With the acceptance set $\mathcal{O}_{\mathrm{acc}} = \{2.36,3\}$ for the Bayesian analysis it leads to a smaller estimated loss. 
On the other hand the optimal Frequentist power of $53.5\%$ with the interval $\{0.58,2.36,3\}$ is significantly lower than for the first testing strategy.
Note, for example in panel c) it might appear that choosing, say the point $1.22$ instead of $0.58$ might be more beneficial for the Frequentist.
However, it needs to be taken into account that the shown distribution is obtained from a worst-case optimization for $\mathcal{O}_{\mathrm{acc}}=\{0.58,2.36,3\}$ and that with appropriate parameters $\mathrm{G}_{\mathrm{acc}}(S\in\{1.22,2.36,3\}|\mathrm{sep}) \geq 30\%$ can be achieved.
}
\label{FIG_FreqVsBayes}
\end{figure}

If the optimization is based on a Frequentist understanding of validity, i.e., achieving at least $70\%$ confidence, it is optimal to use $12$ of the $13$ possible measurement runs and distribute them equally among the three settings as: $n_1 = 4$, $n_2 = 4$ and $n_3 = 4$ (denoted as $[4~4~4]$ for brevity).
Fig.~\ref{FIG_FreqVsBayes}a) illustrates the resulting probability distribution $\mathrm{P}(S|\mathrm{ent})$ and the worst case distribution $\mathrm{P}(S|\mathrm{sep})$ obtained via the discussed optimization, where the best acceptance set turns out to be $\mathcal{O}_{\mathrm{acc}} = \{0,1,2.25,3\}$.
The probability $\mathrm{P}(S \in \mathcal{O}_{\mathrm{acc}}|\mathrm{sep})$ is only $29.8\%$ and thus the confidence remains above $70\%$ while at the same time the power $r = \mathrm{P}(S \in \mathcal{O}_{\mathrm{acc}}|\mathrm{ent})$ is maximized to $66.4\%$.
Estimating also the Bayesian loss for this acceptance interval yields a value of $0.137$.
For comparison, Fig.~\ref{FIG_FreqVsBayes}b) illustrates the estimated posterior distribution.
From the Bayesian perspective, the acceptance set has to be chosen as $\{2.25,3\}$ which leads to a minimal level of acceptance of $76.8\%$ and a loss of $0.149$.
The corresponding power would be $65.6\%$.

This example illustrates two seemingly peculiar consequences of the need to use a worst case estimation.
The first concerns the benefit of using a randomized test.
To maximize the power of the test, the outcomes 0 and 1 are included in the acceptance set.
As seen in Fig.~\ref{FIG_FreqVsBayes} these are not valuable points since they have a significantly higher likelihood given a separable state than an entangled state.
In principle, it would be much more beneficial to add a point with a better likelihood ratio, but as the optimization shows in this case the confidence would drop below $70\%$.
This problem is resolved by the Neymark-Pearson lemma~\cite{Lehmann2005}, which proves that the optimal strategy consists in simply choosing the points with the best likelihood ratios (after corresponding optimization) and make a probabilistic conclusion whenever results with the lowest ratio are obtained.
In this manner the confidence level of $70\%$ could be reached exactly while maximizing the power of the test even further.
Second, the loss of the interval $\{0,1,2.25,3\}$ is estimated to be smaller than the loss of the interval $\{2.25,3.0\}$, while still the latter is chosen as the Bayesian optimum.
This discrepancy is a consequence of the fact that different worst case estimates are chosen to find the acceptance sets and to calculate the loss, as explained further in Methods (Estimating the Probability Distributions).
These loss values are only used to compare the Bayesian performance given different distributions of measurement settings.

Fig.~\ref{FIG_FreqVsBayes}d) shows the posterior distribution for the optimal case according to the Bayesian approach, i.e.~achieving at least $70\%$ level of acceptance with as little expected loss as possible.
It turns out that in this case it is better to use only 11 of the 13 possible measurements and distribute them as $[5~3~3]$ among the three settings.
The requirement of $q_{\mathrm{B}} \geq 70\%$ leads to the choice of acceptance set $\mathcal{O}_{\mathrm{acc}} = \{2.36,3\}$ leading to a loss of $0.146$, which is smaller than the value of $0.149$ obtained with the distribution $[4~4~4]$.
The corresponding power would be $50.4\%$.
Conversely, as illustrated in Fig.~\ref{FIG_FreqVsBayes}c), with $[4~4~4]$ only a power of $53.5\%$ with a level of confidence of $71.1\%$ (and a loss of $0.1603$) can be reached from the point of view of the Frequentist approach making that choice suboptimal as compared with the best choice presented above.
Note, that also here a randomized test would yield better results.

It might seem strange that in both cases a smaller than maximal number of copies is deemed optimal, but this is mainly a consequence of the strong discreteness of the outcome space. 
Indeed the maximal number of $13$ measurements might provide maximal information about the state.
However, the ability to harness this information within the given bounds on the levels of confidence and acceptance is restricted by the need to choose the acceptance and rejection sets from a small number of outcomes.
It can thus happen that for a smaller number of measurements, the probabilities distribute more favorably such that the relevant quantifiers can reach better values.

Finally, it is important to keep in mind that although we subsume the level of confidence and level of acceptance by the term validity, they quantify different figures of merit which correspond to different practical interests.
While the Frequentist is only interested in an average correctness of the conclusion of the certification method, the Bayesian wants to quantify the correctness of the conclusion on a case by case basis (and becomes dependent on the prior).
Note, for example, that with the distribution $[4~4~4]$ although it is much more probable for a separable state to yield a value $1$ than for an entangled state (see Fig.~\ref{FIG_FreqVsBayes}a), given that result the Frequentist would still certify entanglement as this decision increases the overall power without violating the significance level.
On the other hand, if, as done here, a worst case optimization of correlations compatible with separability is used, the Frequentist approach requires prior assumptions only to estimate the power (by modelling the space of entangled states) and can provide an assertion of the confidence which is completely independent of all priors.

\section*{Discussion}
\label{SEC_CONCLUSIONS}

In entanglement certification it is usually of interest to achieve a sufficiently high validity, i.e.~be sufficiently convinced that the certification does not falsely conclude entanglement.
Here we showed explicitly that for small data sets it is necessary to consider not only $\mathrm{P}(Q|\mathrm{sep})$, i.e.~the probability of a separable state to yield particular values of the witness $Q$, but also $\mathrm{P}(Q|\mathrm{ent})$ of entangled states.
Such proper design of an experiment is necessary if we are to make robust conclusions in the presence of finite statistics and allows to define and evaluate the validity and efficiency of the method.
Only with these two quantifiers it becomes possible to optimize the usage of the available experimental resources.

We have introduced a universal procedure that allows the extension of any entanglement witness to the domain of finite statistics.
It is based on an analytical calculation of probability distributions over possible outcomes when estimating correlation functions experimentally with finite resources.
Based on these probability distributions we recalled the measures of validity and efficiency, applicable to any entanglement detection scheme, from the points of view of both Frequentist and Bayesian interpretation.
Illustrative examples were provided based on linear as well as nonlinear witnesses and broad families of states.

The methods introduced here are directly applicable to raw data and should be especially helpful for a resource-efficient estimation of the performance of a known apparatus subject to variations of external parameters or in multipartite experiments with rare detection events, e.g., multi-photon setups based on coincidence clicks. 
For example, our scheme could be employed to quickly certify the quality of a large quantum processor before a time-consuming computation task.

\section*{Methods}

\subsection*{Certifying Entanglement}
Linear witnesses $W$ admit the following expansion in terms of the Pauli basis
\begin{align}
    \mathrm{W} = \sum_{j} \alpha_j \sigma_j.
\end{align}
The expectation value $E_{\mathrm{W}} \equiv \langle \mathrm{W} \rangle$ given a particular state $\rho$, is then a linear function of the correlations $T_j$ given by
\begin{align}
    E_{\mathrm{W}} = \mathrm{Tr}(\mathrm{W} \rho) = \sum_{j} \alpha_j T_j.
    \label{EQ_WITN_LIN}
\end{align}
Measuring each correlation and computing the expectation value of the witness function provides information about the separability of the state.
Constructions of such functions are tailored to a specific entangled state for which $E_{\mathrm{W}} < 0$, whereas the expectation value on separable states is never negative.
For example, a linear witness detecting entanglement of a three-qubit GHZ state could be chosen as
\begin{equation}
\mathrm{W} = \sigma_y \otimes \sigma_y \otimes \sigma_x - \sigma_x \otimes \sigma_x \otimes \sigma_x + 1,
\label{linear_witness_GHZ}
\end{equation}
where the constant ensures that the witness is non-negative on separable states and negative on the GHZ state $(\ket{000} + \ket{111})/\sqrt{2}$, written in the standard $\sigma_z$ basis.

Entanglement in state $\rho$ can also be revealed via nonlinear witnesses.
As an example we consider the scheme introduced in Ref.~\cite{Badziag}.
Consider the so-called correlation length $S$ defined by
\begin{align}
    S \equiv \sum^{3^N}_{j=1} T_j^2,
    \label{EQ_WITN_NON}
\end{align}
which is the sum of all \emph{full} correlations squared. 
Accordingly, the sum goes only up to $3^N$ (the marginal correlations are not included).
It has been proven~\cite{HJ1,HJ2,HJ3,Minh1,Minh2,Knips2016} that a value of $S > 1$ certifies entanglement.
Since $S$ sums up only non-negative terms, entanglement is verified as soon as a subset of $M$ correlations violates the bound, i.e., $S_M>1$.
We omit the subscript whenever the number of measured settings is clear.

\subsection*{Witnesses in Finite Statistics}
\label{Methods_prob}

Since the product of qubit measurement results is either equal to $+1$ or $-1$, the probability that in $n$ trials we obtain $n_+$ products equal to $+1$ is given by the binomial distribution:
\begin{equation}
\mathrm{p}_i(n_+) = \begin{pmatrix}
    n \\
    n_+
\end{pmatrix} 
\left( \frac{1 + T_i}{2} \right)^{n_+} \left( \frac{1 - T_i}{2} \right)^{n_-},
\end{equation}
where $(1 \pm T_i)/2$ is the probability to obtain the product equal to $\pm 1$ in a single trial, estimated from the ideal quantum mechanical prediction. All random variables are assumed to be independent and identically distributed.
Using Eq.~(\ref{EQ_TAU}) the correlation $\tau_i$ is also binomially distributed as
\begin{equation}
\mathrm{P}(\tau_i) = \mathrm{p}_i\left(\frac{n}{2}(1+T_i)\right).
\end{equation}
Note that due to the finiteness of $n$ also $\tau_i$ can take on a finite set of values $\{ -1, -1 + \frac{2}{n}, \dots, 1 - \frac{2}{n}, 1\}$.

In the case of nonlinear witnesses, these correlations have to be further processed. We illustrate this process using the quadratic witness $S$ since it can be easily generalised to any powers.
In order to write the probability distribution of different values of $S$ we first note that
\begin{equation}
\mathrm{P}(\tau_i^2) = \mathrm{P}(\tau_i) + \mathrm{P}(- \tau_i) \quad \textrm{for} \quad \tau_i \ne 0,
\end{equation}
because both values $\pm \tau_i$ give the same square.
In the special case of $\tau_i = 0$, the probability $\mathrm{P}(\tau_i^2)$ is just given by $\mathrm{P}(\tau_i = 0)$.
Since $S$ is defined as the sum of squares, a particular value of $S$ can be realised in many ways,
e.g. the value $S = 1$ can be achieved if any one $\tau_i^2 = 1$ and all the other squared correlations are zero, or when all $\tau_i^2 = 1/M$, etc.
The same reasoning applies to the linear case. Assuming that all the squared correlations $\tau_i^2$ are independently distributed, the values of $S$ satisfy:
\begin{equation}
\mathrm{P}(S) = \sum_{\tau_1^2 + \dots + \tau_M^2 = S} \mathrm{P}(\tau_1^2) \cdots \mathrm{P}(\tau_M^2).
\end{equation}
These distributions were used in the Results section to derive mean values and variances of the introduced quantities.

\subsection*{Estimating the Probability Distributions}

For a simple hypothesis test, determining validity and efficiency requires the calculation of the two likelihoods $\mathrm{P}(Q|\mathrm{sep})$ and $\mathrm{P}(Q|\mathrm{ent})$.
In practice, however, the parameter space is more complicated and one has to consider the full subspaces $\mathcal{H}_\mathrm{sep}$ and $\mathcal{H}_\mathrm{ent}$ of separable and entangled states.
These spaces would be modelled via distributions $\mathrm{P}(\rho_j|\mathrm{sep})$ and $\mathrm{P}(\rho_j|\mathrm{ent})$ of separable and entangled states $\rho_j$ and each of those states would give rise to certain likelihood of outcomes $\mathrm{P}(Q|\rho_j)$.
Although one option would be to expand the discussion towards so called composite hypothesis tests, the simple scenario can be recovered by reducing these state spaces to single parameters (‘sep’ and ‘ent’).
This is done by considering the average distributions of outcomes given a separable or entangled state with
\begin{align}
\mathrm{P}(Q|\mathrm{sep}) = \sum_j \mathrm{P}(Q|\rho_j) \mathrm{P}(\rho_j|\mathrm{sep}), \\
\mathrm{P}(Q|\mathrm{ent}) = \sum_j \mathrm{P}(Q|\rho_j) \mathrm{P}(\rho_j|\mathrm{ent}).
\end{align}
Unfortunately, evaluating such expressions is notoriously daunting~\cite{bergeSCMCsampling}.

In some cases it is possible to avoid the modelling of the state space and instead consider a worst-case distribution of values $Q$.
For $\mathrm{P}(Q|\mathrm{sep})$ this distribution can either be calculated point-wise for each $Q$, denoted as $\mathrm{G}_Q$, or for the whole acceptance interval $\mathcal{O}_{\mathrm{acc}}$, denoted as $\mathrm{G}_{\mathrm{acc}}$.
These distributions have to satisfy
\begin{align}
    \forall_{\rho \in \mathcal{H}_\mathrm{sep}}\, \forall_{Q \in \mathcal{O}_{\mathrm{acc}}} \quad  \mathrm{G}_Q(Q) &\geq \mathrm{P}(Q|\rho)
     \quad \text{and} \\
    \forall_{\rho \in \mathcal{H}_\mathrm{sep}}\ \quad  \mathrm{G}_{\mathrm{acc}}(Q \in \mathcal{O}_{\mathrm{acc}}) &\geq \mathrm{P}(Q \in \mathcal{O}_{\mathrm{acc}}|\rho),
\end{align}
where $\mathrm{G}_Q$ is stronger in the sense that it satisfies the condition for $\mathrm{G}_{\mathrm{acc}}$ as well, i.e.~$\mathrm{G}_{\mathrm{acc}}(Q \in \mathcal{O}_{\mathrm{acc}}) \leq \sum_{Q \in \mathcal{O}_{\mathrm{acc}}} \mathrm{G}_Q(Q)$.
The strategy to find the upper bounds is based on finding a set of $M$ correlations $T_j$ corresponding to $\tilde{\rho}$ such that $\mathrm{P}(Q|\tilde{\rho})$ satisfies the criterion in question.
Note that $\tilde{\rho}$ might not even be a proper density matrix which is why we refer to such correlations as compatible with separability.
Furthermore, in the case of $\mathrm{G}_Q$ the value at each $Q$ can correspond to a different $\tilde{\rho}$.

$\mathrm{G}_{\mathrm{acc}}$ can be used straightforwardly to choose Frequentist acceptance sets by modifying condition (\ref{eq.freq.valid}) to
\begin{align}
    \sum_{Q \in \mathcal{O}_{\mathrm{acc}}} \mathrm{P}(Q|\mathrm{sep}) \leq \sum_{Q \in \mathcal{O}_{\mathrm{acc}}} \mathrm{G}_{\mathrm{acc}}(Q|\mathrm{sep}) \leq 1 - q_{\mathrm{F}}.
\end{align}
If the criterion is satisfied for $\mathrm{G}_{\mathrm{acc}}$ it will also be satisfied for the actual distribution $\mathrm{P}(Q|\mathrm{sep})$.
Thus, the estimation can only lead to a stricter choice of the acceptance set and ensures that the minimal level of significance is attained.

In the Bayesian case for each outcome $Q$ the minimal validity $q_{\mathrm{B}}$ is assured as in Eq.~(\ref{eq.bayes.valid}), by providing a lower bound on $\mathrm{P}(\mathrm{ent}|Q)$ using $\mathrm{G}_Q$ as
\begin{align}
    \mathrm{P}(\mathrm{ent}|Q) = \frac{\mathrm{P}(Q|\mathrm{ent}) \mathrm{P}(\mathrm{ent})}{\mathrm{P}(Q|\mathrm{sep}) \mathrm{P}(\mathrm{sep}) + \mathrm{P}(Q|\mathrm{ent}) \mathrm{P}(\mathrm{ent})} \geq \frac{\mathrm{P}(Q|\mathrm{ent}) \mathrm{P}(\mathrm{ent})}{\mathrm{G}_Q(Q) \mathrm{P}(\mathrm{sep}) + \mathrm{P}(Q|\mathrm{ent}) \mathrm{P}(\mathrm{ent})} \geq q_{\mathrm{B}}.
\end{align}
To estimate the expected loss, however, it is sufficient to employ the weaker worst case estimate $\mathrm{G}_{\mathrm{acc}}$ since the scenario is considered a priori, i.e.~before a particular outcome $Q$ is known.
It is thus sufficient to only maximize the probability in the whole interval $\mathcal{O}_{\mathrm{acc}}$, independently of how $\mathrm{G}_{\mathrm{acc}}(Q)$ is related to $\mathrm{P}(Q|\mathrm{sep})$ point-wise.
The loss (\ref{eq.bayes.loss}) is thus estimated as
\begin{align}
    L &= q_{\mathrm{B}} \mathrm{P}(Q \in \mathcal{O}_{\mathrm{acc}}|\mathrm{sep}) \mathrm{P}(\mathrm{sep}) + (1-q_{\mathrm{B}}) \mathrm{P}(Q \in \mathcal{O}_{\mathrm{rej}}|\mathrm{ent}) \mathrm{P}(\mathrm{ent}) \nonumber \\
    &\leq q_{\mathrm{B}} \mathrm{G}_{\mathrm{acc}}(Q \in \mathcal{O}_{\mathrm{acc}}) \mathrm{P}(\mathrm{sep}) + (1-q_{\mathrm{B}}) \mathrm{P}(Q \in \mathcal{O}_{\mathrm{rej}}|\mathrm{ent}) \mathrm{P}(\mathrm{ent}).
    \label{eq.bayes.loss.est}
\end{align}
Since different worst case estimates are employed, i.e.~$\mathrm{G}_Q$ to find the acceptance interval and $\mathrm{G}_{\mathrm{acc}}$ to estimate the loss, it is no longer the case that the optimal choice of acceptance intervals minimizes the loss.
However, the worst case estimate of the loss with $\mathrm{G}_{\mathrm{acc}}$ is still useful to compare scenarios with different number of copies and distributions of copies over settings, for each of which $\mathrm{G}_Q$ has been used as a rule to find the acceptance sets.

In analogy to $\mathrm{P}(Q|\mathrm{sep})$ in principle both the Frequentist and the Bayesian approach could proceed given a lower bound on $\mathrm{P}(Q|\mathrm{ent})$.
However, this is completely unfeasible for several reasons.
 For example, for the non-linear witness $S$: (i) entangled states can easily yield very small values of $S$ due to statistical fluctuations,
(ii) some entangled states have small (or even vanishing) $S^\infty$~\cite{Kaszlikowski2008,Laskowski2012,Schwemmer2015NoCorr,Tran2017}, 
(iii) there are bad choices of measurement settings, e.g. ones that give $S^{\infty}$ close to 1.
Thus, $\mathrm{P}(S|\mathrm{ent})$ is obtained by modelling a certain state distribution $\mathrm{P}(\rho_j|\mathrm{ent})$, which fits the currently employed apparatus for state generation, as illustrated in Results (Optimising Entanglement Certification Strategies). Similar problems appear for linear witnesses.
Note, that the probability distributions $\mathrm{P}(\rho_j|\mathrm{sep})$ and $\mathrm{P}(\rho_j|\mathrm{ent})$ do not describe a statistical uncertainty of the form encountered in mixed states.
Instead, they capture prior expectations as to what separable or entangled states are to be submitted to the certification procedure.
In other words, due to nonlinearity of the function $\mathrm{P}(Q|\rho_j)$ it does not hold that $\mathrm{P}(Q|\mathrm{sep}) = \mathrm{P}(Q|\bar{\rho}_\mathrm{sep})$ with a mean separable state $\bar{\rho}_\mathrm{sep} = \sum_j \rho_j \mathrm{P}(\rho_j|\mathrm{sep})$ and similarly for $\mathrm{P}(Q|\mathrm{ent})$.

\subsection*{Correlations compatible with separability}
\label{APP_1M}

We have written an optimization algorithm in order to find the set of correlation functions that maximise the probabilities in the acceptance sets while being compatible with separability, i.e.~the correlations do not violate the witness in the limit of infinite sample size.
It verifies that for equal distribution of copies over the settings, i.e.~each $n_i=n$ for each $i=1,\dots,M$, and sufficiently high bound on the linear witness $\mathcal{C}$ ($\mathcal{B}$ for the non-linear witness) all of the correlations can be taken as $T_i=1/M$ ($T_i^2=1/M$).
The algorithm can be also applied to model probability distribution over $Q$ for separable states and arbitrary distributions of copies as well as more complicated disjoint acceptance sets, see Results (Optimising Entanglement Certification Strategies).

The basic algorithm works as follows.
We compute $\mathrm{P}(Q)$ and the cumulative probability function.
They are parameterized with to-be-determined correlations $\lbrace T_1,T_2,...,T_M \rbrace$ as well as the bound defining the acceptance set. 
The heart of the computation is the numerical maximization of $\sum_{Q \in \mathcal{O}_{\mathrm{acc}}} \mathrm{P}(Q) (\mathrm{P}(Q) \forall_{Q \in \mathcal{O}_{\mathrm{acc}}})$ , for the fixed value of bound, with respect to $\lbrace T_1,T_2,...,T_M \rbrace$. 
We constrain the set of possible correlations to those that yield $W^{\infty} \geq 0$ for the linear and $S^{\infty} \leq 1$ for the non-linear witness, with each $T_i \in [-1,1]$.
Two widely available out-of-the-box methods were used for finding the maximum: simulated annealing and Nelder-Mead algorithm ~\cite{Mathematica, scipy}. 
Both gave similar results with the Nelder-Mead method being more precise in the considered problem. 
The output of the algorithm is a list of $\lbrace T_1,T_2,...,T_M \rbrace$ that maximizes $\sum_{Q \in \mathcal{O}_{\mathrm{acc}}} \mathrm{P}(Q) (\mathrm{P}(Q) \forall_{Q \in \mathcal{O}_{\mathrm{acc}}})$ for the given bound, or any other acceptance set, under the stated constraints.

\subsection*{Effects of different types of noise}
\label{APP_NOISE}

\begin{figure}[!t]
\centering
\includegraphics[width=0.6\textwidth]{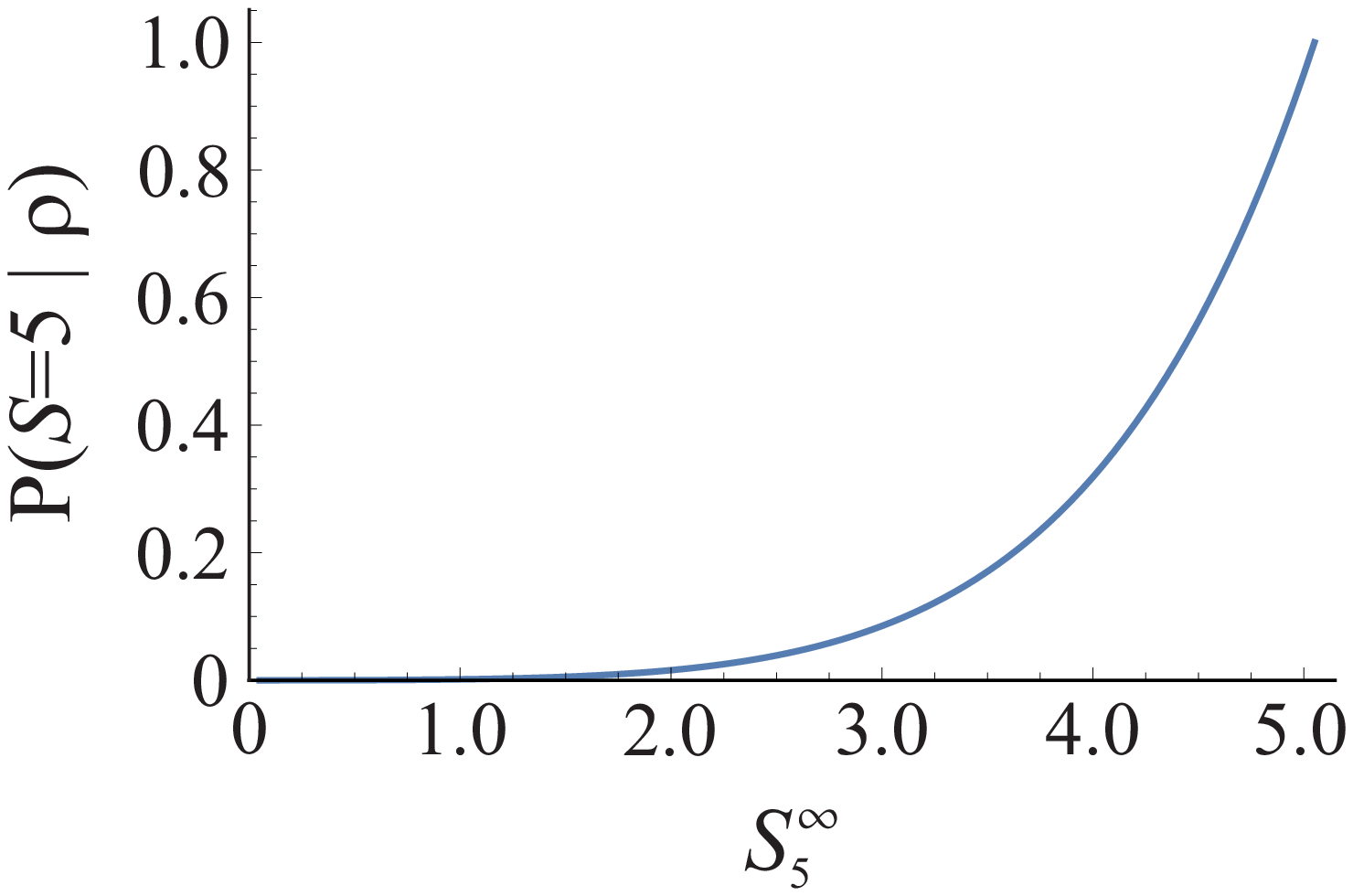}
\caption{
\textbf{The efficiency of entanglement detection for different noises $\rho_n$ and their admixtures $p$ in Eq.~(\ref{EQ_EX_STATE}).}
The plot shows the probability of measuring $S = 5$ given $\rho$ (vertical axis) as a function of $S^\infty_5$ (horizontal axis).
The plotted curve assumes white noise and is given analytically as $[(1 + 6 p^2 + p^4) / 8]^5$.
The data for the other types of noise, i.e., pure and mixed noises according to Eqs.~(\ref{eq:haarrandompure}) and (\ref{eq:haarrandommixed}), overlap with the presented one.
Thus, different types of unbiased noises are not relevant in the considered criteria.
}
\label{FIG_noise}
\end{figure}

We analyse here the effects due to different forms of noise in Eq.~(\ref{EQ_EX_STATE}) on the efficiency of the entanglement certification.
In particular, we consider the following unbiased noises
\begin{align}
    \rho_{\mathrm{wn}} &=\openone/2^N, \quad \text{white noise} \\
    \rho_{\mathrm{rp}} &=|\psi_{\mathrm{r}}\rangle \langle \psi_{\mathrm{r}}|,  \quad \text{Haar random pure state noise} \label{eq:haarrandompure} \\
    \rho_{\mathrm{wn}} &=\rho_{\mathrm{r}}, \quad \text{random mixed states}, \label{eq:haarrandommixed}
\end{align}
where Haar randomness refers to the sampling of states according to the rotationally invariant measure. 
Data for the random states was computed numerically by scanning $p$ in the range $[0.5,1]$ with steps of $0.01$, sampling 1000 random four-qubit states for each $p$, choosing the five biggest correlations, computing $S^{\infty}_5$ and finally the probability of $S = 5$ in finite statistics.
In this way, practically all the values of $S^{\infty}_5$ were scanned. These calculations have shown that the probability of success does not depend on the kind of noise and it is described solely by the value of $p$. The probability of success $\mathrm{P}(S=5|\rho)$ for a continuous choice of $p$ is presented in Fig.~\ref{FIG_noise}. 

\section*{Data availability}

All data supporting Figures are given in the Supplementary Materials.

\section*{Acknowledgments}
We thank Borivoje Daki\'c for useful discussions.
This research was supported by the DFG (Germany) and NCN (Poland) within the joint funding initiative ‘Beethoven2’ (2016/23/G/ST2/04273, 381445721), by the DFG under Germany’s Excellence Strategy EXC-2111-390814868 and by the BMBF project QuKuK (16KIS1621).
PC is supported by the National Science Centre (NCN, Poland) within the Preludium Bis project (Grant No. 2021/43/O/ST2/02679). WL acknowledges partial support from the Foundation for Polish Science (IRAP project ICTQT, Contract No. 2018/MAB/5, co-financed by EU via Smart Growth Operational Programme).
TP is supported by the Polish National Agency for Academic Exchange NAWA Project No. PPN/PPO/2018/1/00007/U/00001.
JD acknowledges support from the PhD program IMPRS-QST.

\bibliographystyle{naturemag}
\bibliography{refs.bib}

\end{document}